\documentclass[12pt]{article}

\usepackage{latexsym,amsmath,amscd,amssymb,amsthm,graphics}
\usepackage{enumerate}
\usepackage[margin=2cm]{geometry}

\usepackage{graphicx}

\usepackage[square,authoryear]{natbib}
\usepackage[colorlinks]{hyperref}
\usepackage{url}
\usepackage{framed}
\usepackage[all]{xy}

\newcommand{\bfi}{\bfseries\itshape}

\makeatletter
\@addtoreset{figure}{section}
\def\thefigure{\thesection.\@arabic\c@figure}
\def\fps@figure{h, t}
\@addtoreset{table}{bsection}
\def\thetable{\thesection.\@arabic\c@table}
\def\fps@table{h, t}
\@addtoreset{equation}{section}

\makeatother

\begin{document}

\title{Euler-Poincar\'{e} equations for anelastic fluid flows
}
\author{Darryl D. Holm
\\Theoretical Division and Center for Nonlinear Studies
\\Los Alamos National
Laboratory, MS B284
\\ Los Alamos, NM 87545}
\date{November 14, 1998. This version November 30, 1998}

\maketitle
\leftline{\Large \bf Abstract}
\bigskip
We show that the ideal (nondissipative) form of the dynamical
equations for the Lipps-Hemler formulation of the anelastic fluid
model follow as Euler-Poincar\'{e} equations, obtained from a
constrained Hamilton's principle expressed in the Eulerian fluid
description. This establishes the mathematical framework for the
following properties of these anelastic equations: the
Kelvin-Noether circulation theorem, conservation of potential
vorticity on fluid parcels, and the Lie-Poisson Hamiltonian
formulation possessing conserved Casimirs, conserved
domain integrated energy and an associated variational principle
satisfied by the equilibrium solutions. We then introduce a modified
set of anelastic equations that represent the mean anelastic motion,
averaged over subgrid scale rapid fluctuations, while preserving the
mathematical properties of the Euler-Poincar\'{e} framework.


\section{Introduction} \label{sec-intro}

The Eulerian formulation of the action principle for an ideal fluid
casts it into a form that is amenable to asymptotic expansions and
thereby facilitates the creation and analysis of approximate fluid
theories. Such an Eulerian action principle results whenever the
general theory of reduction by symmetry groups is applied to
Lagrangian systems, thereby yielding {\bfi Euler--Poincar\'e
equations}, the Lagrangian analog of Lie-Poisson Hamiltonian
equations, \cite{MR[1994]}. This Euler--Poincar\'e setting provides a
shared mathematical structure for many problems in geophysical fluid
dynamics (GFD), with several benefits, both immediate (such as a
systematic approach to hierarchical modeling and versions of Kelvin's
circulation theorem for these models) and longer term (e.g.,
structured multisymplectic integration algorithms).

For example, by using the Euler--Poincar\'e approach,  \cite{HMR[1998a],HMR[1998b]},  find that the action
principles of a variety of incompressible fluid models for GFD are
related by different levels of truncation of asymptotic expansions and
velocity-pressure decompositions, as applied in Hamilton's principle
for the unapproximated Euler equations of rotating stratified ideal
incompressible fluid dynamics. This sequence of GFD models includes
the Euler equations themselves, followed by their approximations,
namely: Euler-Boussinesq equations (EB), primitive equations (PE),
Hamiltonian balance equations (HBE), and generalized Lagrangian mean
(GLM) equations. It also includes rotating shallow water equations
(RSW), semigeostrophic equations (SG), and quasigeostrophic equations
(QG). Thus, asymptotic expansions and velocity-pressure
decompositions of Hamilton's principle for the Euler equations
describing the motion of a rotating stratified ideal incompressible
fluid are used in \cite{HMR[1998a],HMR[1998b]} to cast the
standard models of GFD into Euler-Poincar\'{e} form and thereby unify
these descriptions and their properties at various levels of
approximation. For related developments and additional structure
preserving approximations constructed from a similar viewpoint, see
\cite{AH[1996]}, \cite{AHN[1998]} and \cite{HZ[1998]}.

Recently, \cite{Bannon[1996]} reexamined the anelastic
approximation for deep fluid convection and proposed an alternative
form of the anelastic equations. This alternative model combines
the results of \cite{DF[1969]} and \cite{LH[1982]} to produce a hybrid theory that
(1) conserves the domain integrated energy; (2) preserves potential
vorticity on fluid parcels; and (3) accurately represents the
acoustic adjustment process in Lamb's problem. The equations for a dry
anelastic compressible fluid (atmosphere) rotating at angular
frequency $\boldsymbol{\Omega}$ under constant vertical gravitational
acceleration $g\hat{\bf z}$ take the following
form \cite{Bannon[1996]}
\begin{eqnarray}
\frac{d\mathbf{u}}{dt} + 2 \boldsymbol{\Omega} \times \mathbf{u}
&=&
-\ \nabla\bigg(\frac{p'}{\rho_s}\bigg)
+ \frac{g\theta'}{\theta_s}\hat{\bf z}
\,,
\label{anelastic-mot-eqn}\\
\nabla\cdot(\rho_s\mathbf{u})&=&0
\,,\label{anelastic-div-eqn}\\
\frac{d(\theta_s+\theta')}{dt}&=&0
\,,\label{entropy-advect-eqn}\\
\frac{\theta'}{\theta_s}
&=&
\frac{p'}{\rho_s g H_{\rho}}
- \frac{\rho'}{\rho_s}
\,,\label{diagnostic-rel}\\
\frac{p'}{p_s}
&=&
\frac{\rho'}{\rho_s} + \frac{T'}{T_s}
\,.\label{gas-eos-eqn}
\end{eqnarray}
In these equations, the fluid velocity is denoted $\mathbf{u}$, the
advective time derivative is
$d/dt=\partial/\partial{t}+\mathbf{u}\cdot\nabla$, and the state
variables for this anelastic motion are: pressure $p$, density
$\rho$, specific entropy
$\theta$, and temperature $T$. These state variables consist of the
sum of the base state (with subscript
$s$) and a dynamic contribution denoted with a prime, as in
\begin{equation}\label{theta-decomp}
\theta(x,y,z,t) = \theta_s(z) + \theta'(x,y,z,t)\,,
\end{equation}
where $\theta'$ is the dynamic contribution to the specific entropy
field. The base state is taken to satisfy
\begin{equation}\label{base-rel}
\frac{dp_s}{dz}=-g\rho_s
\,,\quad
p_s=\rho_s R T_s
\,,\quad
C_p\,\theta_s\frac{d\pi_s}{dz}=-\,g
\,,
\end{equation}
where $\pi_s=T_s/\theta_s$. The constants $R$ and $C_p$ are the
ideal gas constant $R$ and the specific heat at constant pressure
$C_p$ for dry air. The density scale height is given
by $1/H_{\rho}=-\rho_s(z)^{-1}d\rho_s/dz$. Given the base state
functions satisfying relations (\ref{base-rel}), as well as the
velocity $\mathbf{u}$ and the dynamic contributions
$p'$ and $\theta'$ at any time, the thermodynamic diagnostic
relations (\ref{diagnostic-rel}) -- (\ref{gas-eos-eqn}) complete the
description. The distinctions between this anelastic model and the
traditional models \cite{DF[1969]}, \cite{LH[1982]} are discussed in
detail by \cite{Bannon[1996]}. For our purposes here, the
important point is that the dynamical equations
(\ref{anelastic-mot-eqn}) - (\ref{entropy-advect-eqn}) agree in the formulations of both \cite{LH[1982]} and  \cite{Bannon[1996]}

In this paper, we show that the dynamical equations
(\ref{anelastic-mot-eqn}) -- (\ref{entropy-advect-eqn}) for the ideal
(nondissipative) anelastic model follow as Euler-Poincar\'{e}
equations, obtained from a constrained Hamilton's principle expressed
in the Eulerian description.  We then introduce a modified
set of anelastic equations that represent the mean anelastic motion,
averaged over subgrid scale rapid fluctuations, while preserving the
mathematical properties of the Euler-Poincar\'{e} framework.
Euler-Poincar\'{e} equations are the Lagrangian analog of Lie-Poisson
Hamiltonian systems \cite{MR[1994]}, \cite{HMR[1998a],HMR[1998b]}. Among other things, the Euler-Poincar\'{e}
formulation of the anelastic fluid equations provides their
Kelvin-Noether circulation theorem. This theorem is the basis for the
conservation of anelastic potential vorticity on fluid parcels. 
Domain-integrated energy is also conserved and the relation of the
Euler--Poincar\'e equations to the Lie-Poisson Hamiltonian
formulation of the anelastic dynamics is given by a Legendre
transformation at the level of the vector fields satisfying the
weighted divergenceless condition (\ref{anelastic-div-eqn}). The
Casimir conservation laws for this Lie-Poisson Hamiltonian
formulation provide a constrained-energy variational principle for
the equilibrium solutions of anelastic dynamics and form a basis for
determining their Lyapunov stability conditions, as done for the
Euler-Boussinesq equations in \cite{AHMR[1986]}.
(The Euler-Boussinesq equations form a special case of the anelastic
model in which the base state is constant.)

In related previous work, a two-dimensional study of the Hamiltonian
structure of the Lipps-Hemler anelastic model was presented by
\cite{SS[1992]}, who also studied
wave-activity conservation laws in the two-dimensional case. A
canonical Hamiltonian formulation of the Lipps-Hemler anelastic model
in three-dimensions was given in \cite{Bernardet[1995]},
Appendix A, for the Lagrangian fluid description of these equations.
Perhaps not unexpectedly, the Lie-Poisson Hamiltonian formulation for
the Eulerian fluid description of these equations agrees the 
canonical formulation of \cite{Bernardet[1995]} and provides
an alternative perspective. Indeed it must, as the Euler-Poincar\'e
theorem \cite{HMR[1998a]} proves for a class of Hamilton's
principles that includes ideal continuum dynamics that the following
four dynamical perspectives of fluid mechanics are equivalent:
Hamilton's principle for the Lagrangian fluid description; the
Euler-Lagrange equations in the Lagrangian fluid description;
Hamilton's principle for the Eulerian fluid description with certain
constrained variations, similar to those for reduced La\-gran\-ge
d'Al\-em\-bert equations; and the Euler--Poincar\'e equations in the
Eulerian fluid description.

The methods of this paper are based on reduction of variational
principles; that is, on Lagrangian reduction. See 
 \cite{Cendra.etal[1987],Cendra.etal[1998a],Cendra.etal[1998b]} and \cite{MS[1993a],MS[1993b]}), who also discuss 
systems with nonholonomic constraints. The latter has been
demonstrated in the work of Bloch, Krishnaprasad, Marsden and
Murray \cite{BKMM[1996]}, who derived the reduced La\-gran\-ge
d'Al\-em\-bert equations for such nonholonomic systems. Coupled with
the methods of the present paper, these techniques for handling
nonholonomic constraints would also be useful, if required, in
continuum systems. In addition, it seems likely that applications
of the techniques of multisymplectic geometry to numerical
integrators associated with multisymplectic reduction will be
exciting developments for the present setting; see \cite{MPS[1998]} for a discussion of this approach.

\paragraph{Organization of the Paper.} In \S\ref{sec-EP} we recall
from \cite{HMR[1998a]} the results of the
Euler-Poincar\'{e} theorem for Lagrangians in continuum
mechanics depending on advected parameters along with their
associated Kelvin--Noether theorem and Lie-Poisson Hamiltonian
formulation. These results establish the mathematical framework into
which we place the dynamical equations for the anelastic model in
\S\ref{sec-anelastic-eqns}. In \S\ref{sec-anelastic-alpha-model} we
introduce a modified set of anelastic equations that represent the
mean anelastic motion, averaged over subgrid scale rapid
fluctuations of amplitude $\alpha$, while preserving the mathematical
properties of the Euler-Poincar\'{e} framework.

\section{Applications of the Euler--Poincar\'{e} Theorem in GFD}
\label{sec-EP}

Here we recall from \cite{HMR[1998a]} the
statements of the Euler--Poincar\'e equations and their associated
Kelvin--Noether theorem in the context of continuum mechanics and
approximate models in geophysical fluid dynamics.

The Euler-Poincar\'{e} equations for a GFD
Lagrangian $L\,[\mathbf{u},D,b\,]$ involve fluid velocity
$\mathbf{u}$, buoyancy (or specific entropy) $b$ and density 
$D$ as functions of three dimensional space with coordinates
$\mathbf{x}$ and time $t$. In vector notation, these equations
are expressed as \cite{HMR[1986],HMR[1998a],HMR[1998b],HMR[1998c]} and \cite{Holm[1996]},
\begin{equation}
\frac{d}{dt} \frac{1}{D} \frac{{\delta} L}{{\delta} \mathbf{u}}
\,+\, \frac{1}{D} \frac{{\delta} L}{{\delta} u^j} \nabla u^j
\,+\, \frac{1}{D} \frac{{\delta} L}{{\delta} b} \nabla b
-  \nabla\frac{{\delta} L}{{\delta} D}=0,
\label{EP-comp1}
\end{equation}
or, equivalently, in ``curl form'' as,
\begin{equation}
\frac{\partial}{\partial t}
\Big(\frac{1}{D} \frac{{\delta} L}{{\delta} \mathbf{u}}\Big)
\,-\, \mathbf{u}\times {\rm curl}
\Big(\frac{1}{D} \frac{{\delta} L}{{\delta} \mathbf{u}}\Big)
\,+\, \nabla\Big(\mathbf{u}\cdot\frac{1}{D}
\frac{{\delta} L}{{\delta} \mathbf{u}}
\,-\, \frac{{\delta} L}{{\delta} D}\Big)
\,+\, \frac{1}{D} \frac{{\delta} L}{{\delta} b} \nabla b
=0\,.
\label{EP-comp2}
\end{equation}
The Euler--Poincar\'e system is completed by including the auxiliary
equations for advection of the buoyancy (or specific entropy) $b$,
\begin{equation}
\frac{\partial{b}}{\partial{t}}+\mathbf{u}\cdot\nabla{b} 
=
0\,,
\label{b-advect}
\end{equation}
and the continuity equation for the density $D$,
\begin{equation}
\frac{\partial{D}}{\partial{t}}
+\nabla\cdot({D}\mathbf{u}) 
=
0\,.
\label{D-cont}
\end{equation}
For incompressible flows, one sets $D=1$ in the continuity equation,
so that $\nabla \cdot \mathbf{u} = 0$. For anelastic flows, one sets
$D=\rho_s(z)$ in the continuity equation with a prescribed stably
stratified reference density profile $\rho_s(z)$, so that
$\nabla\cdot(\rho_s(z)\mathbf{u})=0$. 

The Euler--Poincar\'e motion equation in either form
(\ref{EP-comp1}) or (\ref{EP-comp2}) results in the {\bfi
Kelvin-Noether circulation theorem},
\begin{equation} \label{KN-theorem-bD}
\frac{d}{dt}\oint_{\gamma_t(\mathbf{u})} \frac{1}{D}\frac{\delta
L}{\delta \mathbf{u}}\cdot d\mathbf{x}
= -\oint_{\gamma_t(\mathbf{u})}
\frac{1}{D}\frac{\delta L}{\delta b}\nabla b \cdot d\mathbf{x}\;,
\end{equation}
where the curve $\gamma_t(\mathbf{u})$ moves with the fluid velocity
$\mathbf{u}$. Then, by Stokes' theorem, the Euler--Poincar\'e
equations generate circulation of the quantity
$D^{-1}{\delta{L}/\delta\mathbf{u}}$ whenever the
gradients $\nabla b$ 
and $\nabla(D^{-1}\delta L/\delta{b})$ are not collinear.

Taking the curl of equation (\ref{EP-comp2}) and using advection of
the buoyancy $b$ and the continuity equation for the density $D$
yields {\bfi conservation of potential vorticity on fluid
parcels}, as expressed by
\begin{equation} \label{pv-cons-EP}
\frac{\partial{q}}{\partial{t}}+\mathbf{u}\cdot\nabla{q} =
0\,,
\quad \hbox{where}\quad
{q}\equiv\frac{1}{D}\nabla{b}\cdot{\rm curl}
\left(\frac{1}{D}\frac{\delta L}{\delta \mathbf{u}}\right).
\end{equation}
Consequently, the following domain integrated quantities are
conserved, for any function $\Phi$,
\begin{equation} \label{EP-Casimirs}
C_{\Phi} = \int d^{\,3}x\ 
D\,\Phi(b,q)\,,
\quad\forall\,\Phi\,.
\end{equation}
The absence of explicit time dependence in the Lagrangian
$L\,[\mathbf{u},D,b\,]$ gives the {\bfi conserved domain integrated
energy}, via Noether's theorem for time translation invariance. This
energy is easily calculated using the {\bfi Legendre transform} to be
\begin{equation} \label{EP-erg}
E\,[\mathbf{u},D,b\,] = \int d^{\,3}x\ 
\Big(\mathbf{u}\cdot
\frac{\delta L}{\delta \mathbf{u}}\Big)
- L\,[\mathbf{u},D,b\,]
\,.
\end{equation}
When the Legendre transform is completed to express
$E\,[\mathbf{u},D,b\,]$ as $H\,[\mathbf{m},D,b\,]$ with
$\mathbf{m}\equiv\delta L/\delta \mathbf{u}$ and 
$\delta{H}/\delta\mathbf{m}=\mathbf{u}$, the Euler--Poincar\'e system
(\ref{EP-comp1})--(\ref{D-cont}) may be expressed in Hamiltonian form
\begin{equation}\label{mu-dot-syst}
\frac{\partial \mu}{\partial t}=\{\mu,H\}\,,
\quad \hbox{with}\quad
\mu\in[\mathbf{m},D,b\,]\,,
\end{equation}
and {\bfi Lie-Poisson bracket} given in Euclidean component form by
\begin{eqnarray}\label{LPB}
&&\hspace{-.5in}
\{F,H\}[\mathbf{m},D,b\,] 
\\
&& =\ - \, \int d^{\,3}x\ \bigg\{
\frac{\delta F}{\delta m_i}
\bigg[\big(\partial_j m_i+m_j \partial_i\,\big)
\frac{\delta H}{\delta m_j}
\ +\ \big(D\partial_i\,\big)\frac{\delta H}{\delta D}
\ -\ \big( b_{,i}\,\big)\frac{\delta H}{\delta b}\bigg]
\nonumber\\
&&\hspace{1in}
+\ \frac{\delta F}{\delta D}\big(\partial_j D\big)
\frac{\delta H}{\delta m_j}
\ +\ \frac{\delta F}{\delta b}
\big( b_{,j}\big)
\frac{\delta H}{\delta m_j}
\bigg\}\,.
\nonumber
\end{eqnarray}
The conserved quantities $C_{\Phi}$ in (\ref{EP-Casimirs}) are
then understood in the {\bfi Lie-Poisson Hamiltonian  formulation}
(\ref{mu-dot-syst}) -- (\ref{LPB}) of the Euler--Poincar\'e system
(\ref{EP-comp1}) -- (\ref{D-cont}) as {\bfi Casimirs} that commute
under the Lie-Poisson bracket (\ref{LPB}) with any functional of
$[\mathbf{m},D,b\,]$. The Casimirs also result via Noether's theorem
from symmetry of the Hamilton's principle for the Euler--Poincar\'e
system under the ``particle relabeling transformations'' that leave
invariant the Lagrangian $L[\mathbf{u},D,b\,]$. From the viewpoint
of Noether's theorem, this particle relabeling symmetry corresponds to
invariance of the Hamilton's principle for the  Euler--Poincar\'e
equations under the transformation from the Lagrangian to the
Eulerian fluid description, by pullback of the right action of the
diffeomorphism group on the configuration space of the Lagrangian
fluid parcel positions and their velocities.
For full mathematical details, consult \cite{MR[1994],HMR[1998a],HMR[1998b],HMR[1998c]}.

The four properties (\ref{KN-theorem-bD})--(\ref{EP-erg}) and the
Lie-Poisson Hamiltonian formulation (\ref{mu-dot-syst}) --
(\ref{LPB}) of the Euler--Poincar\'e equation (\ref{EP-comp1}) and
its auxiliary equations (\ref{b-advect}) and (\ref{D-cont}) are
desirable elements of approximate models for applications in
geophysical fluid dynamics expressed in the variables
$[\mathbf{u},D,b\,]$. Thus, the Euler--Poincar\'e theory offers a
unified framework in which to derive approximate GFD models that
possess these properties: the Kelvin-Noether circulation
theorem, conservation of potential vorticity on fluid parcels, and the
Lie-Poisson Hamiltonian formulation with its associated conserved
Casimirs and conserved domain integrated energy. Previous 
work of \cite{HMR[1998a],HMR[1998b],HMR[1998c]} has
shown that many useful GFD approximations may be formulated as
Euler--Poincar\'e equations, whose shared properties thus follow from
this underlying common framework. The aim of the next section of this
paper is to cast the dynamical anelastic equations
(\ref{anelastic-mot-eqn}) -- (\ref{entropy-advect-eqn}) into the
Euler--Poincar\'e framework, as well.

\section{The dynamical anelastic equations are Euler--Poincar\'e}
\label{sec-anelastic-eqns}

\subsection*{The Lagrangian} In the Eulerian fluid representation,
we consider Hamilton's principle for fluid motion in a three
dimensional domain with action functional ${\cal S}=\int\,dt\, L$
and Lagrangian $L[\mathbf{u},D,b\,]$ given by
\begin{equation}
L = \int d^{\,3}x\ D
\bigg(\frac{1}{2} |\mathbf{u}|^2
+ \mathbf{u}\cdot\mathbf{R}(\mathbf{x}) - gz
- C_p\pi_s(z)b\bigg)
+ p'\bigg(1-\frac{D}{\rho_s(z)}\bigg)\,,
\label{lag-anelastic}
\end{equation}
where $D$ is the mass density and $\mathbf{R}(\mathbf{x})$,
$\pi_s(z)$, and $\rho_s(z)$ are given functions of their arguments.
This Lagrangian produces the following variations at fixed
$\mathbf{x}$ and $t$
\begin{eqnarray}
&&\frac{1}{D}\frac{{\delta} L}{{\delta} \mathbf{u}}
= \mathbf{u}+ \mathbf{R}(\mathbf{x})\,,
\quad
\frac{{\delta} L}{{\delta} b}
= - C_p\pi_s(z)\,,
\quad
\frac{{\delta} L}{{\delta} p'} = 1-\frac{D}{\rho_s(z)}\,,
\nonumber \\
&&\frac{{\delta} L}{{\delta} D}
= \frac{1}{2} |\mathbf{u}|^2
+ \mathbf{u}\cdot\mathbf{R}(\mathbf{x}) - gz
- C_p\pi_s(z)b - \frac{p'}{\rho_s(z)}\,.
\hspace{.25in}
\label{vds-1}
\end{eqnarray}
Hence, from the Euclidean component formula (\ref{EP-comp1}) for
Hamilton principles of this type, we find the motion equation for
such a fluid in three dimensions,
\begin{equation}
\frac{d\mathbf{u}}{dt} 
- \mathbf{u} \times {\rm curl} \mathbf{R}
+\ \nabla\bigg(\frac{p'}{\rho_s}\bigg)
+ \bigg(g + C_p b \frac{d\pi_s}{dz}\bigg)\hat{\bf z}
=0
\,,
\label{EP-mot}
\end{equation}
where ${\rm curl}\,\mathbf{R}=2\boldsymbol{\Omega}(\mathbf{x})$ is the
Coriolis parameter (i.e., twice the local angular rotation
frequency). We use equation (\ref{base-rel}) to rewrite the last  
term in parentheses as
\begin{equation}
g + C_p b \frac{d\pi_s}{dz} 
= g\bigg(1-\frac{b}{\theta_s}\bigg)
= - g \frac{\theta'(\mathbf{x},t)}{\theta_s(z)}\,,
\label{reln1}
\end{equation}
in which we identify $b$ as the total specific entropy, 
\begin{equation}\label{theta-decomp-b}
b = \theta(\mathbf{x},t) = \theta_s(z) + \theta'(\mathbf{x},t)\,,
\end{equation}
since each satisfies the scalar advection relation
(\ref{entropy-advect-eqn}), cf. (\ref{b-advect}). Hence, from
(\ref{EP-mot}) and (\ref{reln1}) we recover the anelastic motion
equation (\ref{anelastic-mot-eqn}), namely,
\begin{equation}
\frac{d\mathbf{u}}{dt} 
- \mathbf{u} \times 2\boldsymbol{\Omega}(\mathbf{x})
+\ \nabla\bigg(\frac{p'}{\rho_s}\bigg)
- \frac{g\theta'}{\theta_s}\hat{\bf z}
=0
\,,
\label{anelastic-mot-eqn1}
\end{equation}
as the
Euler--Poincar\'e equation for the Lagrangian (\ref{lag-anelastic}).
Finally, we substitute the contraint $D=\rho_s(z)$ obtained from
stationarity of the Lagrangian (\ref{lag-anelastic}) with respect to
variations in $p'$ into the continuity equation (\ref{D-cont}) to
find the anelastic divergence condition
$\nabla\cdot\rho_s\mathbf{u}=0$, i.e.,
equation (\ref{anelastic-div-eqn}). Preservation of this condition
determines the dynamic pressure contribution, $p'$, by solving the
elliptic equation obtained by taking the divergence of
the anelastic motion equation (\ref{anelastic-mot-eqn1}) after
multiplying it by the base density $\rho_s(z)$, 
\begin{equation}
- \Delta p\,' = g \frac{\partial\rho\,'}{\partial z}
+
{\rm div}\,(\hbox{nonlinearity})
\,.
\label{anelastic-mot-eqn1}
\end{equation}
The boundary
condition for this elliptic equation is obtained from the normal
component of the anelastic motion equation (\ref{anelastic-mot-eqn1})
evaluated on the boundary and using the boundary condition for the
velocity, e.g., that it has no normal component at the boundary,
which yields a Neumann boundary condition for obtaining the pressure.
\begin{equation}
-\,\frac{\partial{p}\,'}{\partial n} = g\rho\,'
(\mathbf{\hat{n}}\cdot\mathbf{\hat{z}})
+
\mathbf{\hat{n}}\cdot(\hbox{nonlinearity})
\,.
\label{anelastic-mot-eqn1}
\end{equation}
See the treatment in \cite{Bernardet[1995]} for a
discussion of alternative velocity boundary conditions for the
Lipps-Hemler anelastic model.

\subsection*{The Kelvin--Noether theorem} From equation
(\ref{KN-theorem-bD}), the Kelvin--Noether circulation theorem
corresponding to the anelastic motion equation
(\ref{anelastic-mot-eqn1}) for an ideal anelastic fluid
in three dimensions is,
\begin{equation}
\frac{d}{dt}\oint_{\gamma_t(\mathbf{u})}(\mathbf{u}
+\mathbf{R})\cdot d\mathbf{x}
= -\oint_{\gamma_t(\mathbf{u})}
C_p\, \theta\, \nabla \pi_s(z)\cdot d\mathbf{x}\;,
\label{KN-theorem-Anel}
\end{equation}
where the curve $\gamma_t(\mathbf{u})$ moves with the anelastic fluid
velocity $\mathbf{u}$. By Stokes' theorem, the anelastic equations
generate circulation of $(\mathbf{u}+\mathbf{R})$ around
$\gamma_t(\mathbf{u})$ whenever the gradient of specific entropy
$\theta$ has a horizontal component. Using advection of $\theta$
and the anelastic divergence condition, one finds conservation of
potential vorticity $q_{\rm Anel}$ on fluid parcels, cf. equation
(\ref{pv-cons-EP}),
\begin{equation} \label{pv-cons-Anel}
\frac{\partial{q}_{\rm Anel}}{\partial{t}}
+\mathbf{u}\cdot\nabla{q}_{\rm Anel} = 0\,,
\quad \hbox{where}\quad
{q}_{\rm Anel} = \frac{1}{\rho_s(z)}
\nabla\theta\cdot{\rm curl}\,(\mathbf{u}+\mathbf{R})\,.
\end{equation}
Consequently, the following domain integrated quantities are
conserved, for any function $\Phi$, cf. equation (\ref{EP-Casimirs}),
\begin{equation} \label{Anel-Casimirs}
C_{\Phi} = \int d^{\,3}x\ 
\rho_s(z)\,\Phi(\theta,q_{\rm Anel})\,,
\quad\forall\,\Phi\,.
\end{equation}
%
\subsection*{Energy conservation, Lie-Poisson Hamiltonian
formulation and nonlinear Lyapunov stability analysis} 

The conserved anelastic energy is easily calculated using the
Legendre transform of the Lagrangian (\ref{lag-anelastic}) to be
\begin{equation} \label{Anel-erg}
E_{Anel} = \int d^{\,3}x\ \rho_s(z)
\bigg(\frac{1}{2} |\mathbf{u}|^2
+ gz
+ C_p\pi_s(z)\theta\bigg)
\,.
\end{equation}
The corresponding Hamiltonian is (with $b=\theta$)
\begin{equation}\label{Anel-LP-Ham}
H_{Anel} = \int d^{\,3}x\ 
\bigg(\frac{1}{2D} |\mathbf{m}-D\mathbf{R}|^2
+ Dgz
+ C_p\pi_s(z)\,D\,b\bigg)
+ p'\bigg(\frac{D}{\rho_s(z)}-1\bigg)\,.
\end{equation}
The Lie-Poisson bracket (\ref{LPB}) now generates the dynamical
anelastic equations (\ref{anelastic-mot-eqn})
-- (\ref{entropy-advect-eqn}) from this Hamiltonian
according to equations (\ref{mu-dot-syst}). 

The canonical Hamiltonian formulation of the Lipps-Hemler dynamics due
to \cite{Bernardet[1995]}, Appendix A, is based on the
Hamiltonian in the Lagrangian fluid description,
\begin{equation} \label{Lag-Anel-Ham}
H_{canon} = \int d^{\,3}a\ 
\bigg(\frac{1}{2} |\boldsymbol{\dot{x}}(\mathbf{a},t)|^2
+ C_p\pi_s(z(\mathbf{a},t))\,b(\mathbf{a})\bigg)
\,.
\end{equation}
Transforming this Hamiltonian to the Eulerian fluid description
yields,
\begin{equation}\label{Anel-LP-Ham2}
H_{Anel} = \int d^{\,3}x\ 
\bigg(\frac{1}{2D} |\mathbf{m}-D\mathbf{R}|^2
+ C_p\pi_s(z)\,D\,b\bigg)
+ p'\bigg(\frac{D}{\rho_s(z)}-1\bigg)\,,
\end{equation}
in which we again impose the anelastic density constraint explicitly,
by using the pressure $p'$ as a Lagrange multiplier.
Applying the Lie-Poisson bracket (\ref{LPB}) with this Hamiltonian
yields the same dynamical anelastic equations as in 
(\ref{anelastic-mot-eqn}) -- (\ref{entropy-advect-eqn}), up to a
redefinition of pressure to incorporate the gravitational
acceleration. Thus, as guaranteed by the Euler-Poincar\'e theorem
\cite{HMR[1998a]} and the general theory of reduction \cite{MR[1994]},
the Lie-Poisson Hamiltonian formulation of the dynamical anelastic
equations presented here in the Eulerian fluid description is
equivalent to the canonical Hamiltonian formulation due to 
\cite{Bernardet[1995]} in the Lagrangian fluid description.

In the Eulerian fluid description we use the Casimir conserved
quantities (\ref{EP-Casimirs}) to find the following {\bfi
variational principle for anelastic equilibrium solutions}: The
equilibrium solutions of the dynamical anelastic equations occur at
critical points of the sum
$H_{\Phi}$, where
\begin{equation} \label{crt-pt}
H_{\Phi} = H_{Anel} + C_{\Phi}
\,,
\end{equation}
and
\begin{equation} \label{casimir-def}
C_{\Phi} = \int d^{\,3}x\ 
D\,\Phi(b,q)\,,
\quad\forall\,\Phi\,,
\quad\hbox{where}\quad
{q}\equiv\frac{1}{D}\nabla{b}\cdot{\rm curl}
\big(\mathbf{m}/D\big).
\end{equation}
Thus, the Casimir conservation laws for this Lie-Poisson Hamiltonian
formulation of the three-dimensional anelastic equations in the
Eulerian fluid description provide a constrained-energy variational
principle for the equilibrium solutions of the dynamical anelastic
equations and form a basis for determining their Lyapunov stability
conditions, as done for the Euler-Boussinesq equations in \cite{AHMR[1986]}. The Euler-Boussinesq equations form a
special case of the dynamical anelastic equations in which the base
state is constant. Consequently, the analysis of the nonlinear
Lyapunov stability conditions for the equilibrium solutions of
the three-dimensional anelastic equations follows a similar
procedure to that performed in \cite{AHMR[1986]} and
produces a similar result, modulo the nonconstant base state.

\subsection*{The pseudo-incompressible approximation (PIA) }

The PIA  of \cite{Durran[1989]} enhances the anelastic equations by allowing the influence of the base specific entropy field $\theta_s(z)$ on the mass balance. 
Additional studies of the PIA and comparisons of its performance with the dynamics of the anelastic and Euler-Boussinesq models appear in~\cite{ND[1994]},~\cite{Lilly[1994]}.
\\

As an Euler-Poincar\'e system, the PIA equations arise from a
modification of the anelastic Lagrangian in equation
(\ref{lag-anelastic-alpha}). Namely, we consider the Lagrangian for
the PIA system given by
\begin{eqnarray}
\ell_{PIA} &=& \int d^{\,3}x\ \bigg[ D
\bigg(\frac{1}{2} |\mathbf{u}|^2
+ \mathbf{u}\cdot\mathbf{R}(\mathbf{x}) - gz
- C_p\pi_s(z)\,\theta\bigg)
\\
&&\hspace{0.5in}
+
\tilde{p}
\underbrace{\,
\Big(
 \rho_s(z)\theta_s(z) - D\,\theta\Big)
}_{\hbox{PIA constraint}}\
\bigg]
\,.
\nonumber\label{lag-pseudo-inc-alpha}
\end{eqnarray}
This Lagrangian slightly modifies the volume constraint in
the anelastic Lagrangian (\ref{lag-anelastic-alpha}). 
The modified PIA constraint on $D$ imposed by the pressure $\tilde{p}$ requires 
\begin{equation} 
D^{\ast}
\equiv
D\,\theta = \rho_s(z)\,\theta_s(z)
\,.
\nonumber\label{constraint-pseudo-inc}
\end{equation}
Hence, the fluid velocity $\mathbf{u}$ must satisfy
a weighted incompressibility relation ({pseudo-incompressibility})
\begin{equation}
\frac{\partial{D}^{\ast}}{\partial{t}}
+\nabla\cdot({D^{\ast}}\mathbf{u}) 
=
0\,,
\quad\Rightarrow\quad
\nabla\cdot\big(\rho_s(z)\,\theta_s(z)\mathbf{u}\big) 
=0
\,,
\label{pseudo-incompressibility}
\end{equation}
obtained via time independence of ${D}^{\ast}$, and the advection equations for specific entropy $\theta$ and density $D$,
\begin{equation}
\frac{\partial\theta}{\partial{t}}+\mathbf{u}\cdot\nabla\theta 
=
0
\,,\qquad
\frac{\partial{D}}{\partial{t}}
+\nabla\cdot({D}\mathbf{u}) 
=
0\,.
\label{theta-D-advect}
\end{equation}
The PIA equations are the same as for the anelastic model
except for the effects of the differently weighted
incompressibility constraint in (\ref{constraint-pseudo-inc}).

\section{The dynamical anelastic-alpha equations are
Euler--Poincar\'e}
\label{sec-anelastic-alpha-model}

\subsection*{The Lagrangian for the anelastic-alpha model} 

Following \cite{HMR[1998a],HMR[1998b]}, we introduce into the dynamical anelastic
equations the effects of averaging over rapid fluctuations whose
amplitude falls below the length scale denoted as $\alpha$, by making
the following modification of the Lagrangian in
equation (\ref{lag-anelastic}) for the anelastic model
\begin{eqnarray}
L &=& \int d^{\,3}x\ \bigg[ D
\bigg(\frac{1}{2} |\mathbf{u}|^2
+ \frac{\alpha^2}{2} |\nabla\mathbf{u}|^2
+ \mathbf{u}\cdot\mathbf{R}(\mathbf{x}) - gz
- C_p\pi_s(z)b\bigg)
\nonumber\\
&&\hspace{1in}
+ p'\bigg(1-\frac{D}{\rho_s(z)}\bigg)\bigg]\,,
\label{lag-anelastic-alpha}
\end{eqnarray}
where $|\nabla\mathbf{u}|^2 \equiv
\mathbf{u}_{,k}\cdot\mathbf{u}_{,k}$ and we denote the other variables
as before, in equation (\ref{lag-anelastic}). This modified Lagrangian
for the anelastic-alpha model produces the following variations at
fixed $\mathbf{x}$ and $t$
\begin{eqnarray}
&&\frac{1}{D}\frac{{\delta} L}{{\delta} \mathbf{u}}
= \mathbf{u} 
- \frac{\alpha^2}{D}\big(D\mathbf{u}_{,k}\big)_{,k}
+ \mathbf{R}(\mathbf{x})\,,
\quad
\frac{{\delta} L}{{\delta} b}
= - C_p\pi_s(z)\,,
\quad
\frac{{\delta} L}{{\delta} p'} = 1-\frac{D}{\rho_s(z)}\,,
\nonumber \\
&&\frac{{\delta} L}{{\delta} D}
= \frac{1}{2} |\mathbf{u}|^2
+ \frac{\alpha^2}{2} |\nabla\mathbf{u}|^2
+ \mathbf{u}\cdot\mathbf{R}(\mathbf{x}) - gz
- C_p\pi_s(z)b - \frac{p'}{\rho_s(z)}\,.
\hspace{.25in}
\label{vds-1}
\end{eqnarray}
Hence, from the Euclidean component formula (\ref{EP-comp1}) for
Hamilton principles of this type, the Euler--Poincar\'e equation for
the Lagrangian (\ref{lag-anelastic-alpha}) is given by 
\begin{equation}
\frac{d\mathbf{v}}{dt} + v_j \nabla u^j
- \mathbf{u} \times {\rm curl}\mathbf{R}
+\ \nabla\bigg(\frac{p'}{\rho_s}
- \frac{1}{2} |\mathbf{u}|^2
- \frac{\alpha^2}{2} |\nabla\mathbf{u}|^2\bigg)
- \frac{g\theta'}{\theta_s}\hat{\bf z}
=0
\,.
\label{anelastic-alpha-mot-eqn1}
\end{equation}
This is the motion equation for the anelastic-alpha
fluid in three dimensions. One should compare this motion
equation with equation (\ref{anelastic-mot-eqn1}) for the standard
anelastic fluid and note that the advective time derivative is still
defined as
$d/dt=\partial/\partial{t} + \mathbf{u}\cdot\nabla$, in terms of the
fluid parcel transport velocity, $\mathbf{u}$. We have used the
following additional notation in equation
(\ref{anelastic-alpha-mot-eqn1}),
\begin{equation}
\mathbf{v} \equiv \mathbf{u} 
- \alpha^2 \tilde\Delta \mathbf{u}\,,
\quad\hbox{and}\quad
\tilde\Delta \mathbf{u} \equiv 
\frac{1}{D}\big(D\mathbf{u}_{,k}\big)_{,k}
\,,
\label{vee-def}
\end{equation}
where the weighted Laplacian operator $\tilde\Delta$ is given by
\begin{equation}
\tilde\Delta =
\frac{1}{\rho_s(z)}
\bigg(\frac{\partial}{\partial x^k}
\rho_s(z)
\frac{\partial}{\partial x^k}
\bigg)
\quad\hbox{for}\quad D=\rho_s(z)
\,,
\label{vee-def}
\end{equation}
and subscript-comma index notation denotes partial spatial
derivatives. The key observation about the anelastic-alpha model is
that its transport velocity $\mathbf{u}$ is smoother relative to its
momentum or circulation velocity $\mathbf{v}$, by inversion of the
Helmholtz operator, $(1-\tilde\Delta)$, which depends upon the
stratification profile of the base state. The curl form of equation
(\ref{anelastic-alpha-mot-eqn1}) is 
\begin{equation}
\frac{\partial\mathbf{v}}{\partial{t}} 
- \mathbf{u} \times {\rm curl}\,(\mathbf{R} + \mathbf{v})
+\ \nabla\bigg(\frac{p'}{\rho_s}
- \frac{1}{2} |\mathbf{u}|^2
- \frac{\alpha^2}{2} |\nabla\mathbf{u}|^2
+ \mathbf{u}\cdot\mathbf{v} \bigg)
- \frac{g\theta'}{\theta_s}\hat{\bf z}
=0
\,.
\label{anelastic-alpha-mot-eqn2}
\end{equation}
As we shall see, the analysis of the anelastic-alpha model follows the
same procedure as for the anaelastic model, with appropriate changes
from $\mathbf{u}$ to $\mathbf{v}$ to account for the effects of the
averaging over the subgrid (sub-$\alpha$-length) scale
rapid fluctuations.

\subsection*{The Kelvin--Noether theorem for the anelastic-alpha
model} 

From equation (\ref{KN-theorem-bD}), the Kelvin--Noether
circulation theorem corresponding to the anelastic motion equation
(\ref{anelastic-mot-eqn1}) for an ideal anelastic-alpha
model in three dimensions is,
\begin{equation}
\frac{d}{dt}\oint_{\gamma_t(\mathbf{u})}(\mathbf{v}
+\mathbf{R})\cdot d\mathbf{x}
= -\oint_{\gamma_t(\mathbf{u})}
C_p\, \theta\, \nabla \pi_s(z)\cdot d\mathbf{x}\;,
\label{KN-theorem-Anel-alpha}
\end{equation}
where the curve $\gamma_t(\mathbf{u})$ follows the fluid parcels
in moving with anelastic-alpha fluid velocity $\mathbf{u}$. Thus, by
Stokes' theorem, the anelastic-alpha equations generate circulation of
$(\mathbf{v}+\mathbf{R})$ around
$\gamma_t(\mathbf{u})$ whenever the gradient of specific entropy
$\theta$ has a horizontal component. Using advection of $\theta$
and the anelastic divergence condition, one finds conservation of
the anelastic-alpha potential vorticity $q_{{\rm A}-\alpha}$ on fluid
parcels, which are transported with the anelastic velocity
$\mathbf{u}$, cf. equation (\ref{pv-cons-EP}),
\begin{equation} \label{pv-cons-Anel-alpha}
\frac{\partial{q}_{{\rm A}-\alpha}}{\partial{t}}
+\mathbf{u}\cdot\nabla{q}_{{\rm A}-\alpha} = 0\,,
\quad \hbox{where}\quad
{q}_{{\rm A}-\alpha} = \frac{1}{\rho_s(z)}
\nabla\theta\cdot{\rm curl}\,(\mathbf{v}+\mathbf{R})\,.
\end{equation}
Consequently, the following domain integrated quantities are
conserved, for any function $\Phi$, cf. equation (\ref{EP-Casimirs}),
\begin{equation} \label{Anel-Casimirs}
C_{\Phi} = \int d^{\,3}x\ 
\rho_s(z)\,\Phi(\theta,q_{{\rm A}-\alpha})\,,
\quad\forall\,\Phi\,.
\end{equation}
%

\subsection*{Energy conservation for the anelastic-alpha model} 

The conserved anelastic-alpha energy is calculated as before using
the Legendre transform of the Lagrangian (\ref{lag-anelastic-alpha})
and found to be
\begin{equation} \label{Anel-alpha-erg}
E_{{\rm A}-\alpha} = \int d^{\,3}x\ \rho_s(z)
\bigg(\frac{1}{2} |\mathbf{u}|^2
+ \frac{\alpha^2}{2} |\nabla\mathbf{u}|^2
+ gz
+ C_p\pi_s(z)\theta\bigg)
\,.
\end{equation}
Thus, the kinetic energy is augmented in the anelastic-alpha model by
a term proportional to the squared amplitude of the velocity shear.
Hence, the anelastic-alpha model costs energy to produce velocity
shear and its solutions will tend to have smoother velocity profiles
than those for the standard anelastic model.

The curl of equation (\ref{anelastic-alpha-mot-eqn2}) and the
anelastic divergence condition $\nabla\cdot\rho_s\mathbf{u}=0$ yield
an equation for anelastic-alpha vortex dynamics,
\begin{equation}
\frac{\partial\mathbf{q}}{\partial{t}} 
+ \mathbf{u}\cdot\nabla\mathbf{q}
= \mathbf{q}\cdot\nabla\mathbf{u}
+
\frac{1}{\rho_s}\nabla\bigg(\frac{\theta'}{\theta_s}\bigg)
\times g\hat{\bf z}
\,,
\quad\hbox{where}\quad
\mathbf{q}\equiv
\frac{1}{\rho_s(z)}{\rm curl}(\mathbf{v}+\mathbf{R})
\,.
\label{anelastic-vortex-eqn}
\end{equation}
The control on the $L^2$ norm $\|\nabla\mathbf{u}\|_2$ afforded by
the conserved energy in equation (\ref{Anel-alpha-erg}) should tend
to moderate the vortex stretching term
$\mathbf{q}\cdot\nabla\mathbf{u}$ in the vortex dynamics equation
(\ref{anelastic-vortex-eqn}) and, thus, produce less violent,
smoother and more coherent turbulent vortex dynamics than occurs for
the standard anelastic equations. Numerical investigation of the
anelastic-alpha system introduced here will be conducted and reported
elsewhere.

\subsection*{Acknowledgments}
\addcontentsline{toc}{section}{Acknowledgments}
We thank Piotr Smolarkiewicz for stimulating this work and providing
helpful comments on an early draft. This work was conducted under the
auspices of the US Department of Energy, Climate Change and
Predictability Program.

\bibliographystyle{new}

\addcontentsline{toc}{section}{References}

\begin{thebibliography}{300}

\footnotesize 

\bibitem[Abarbanel et al.(1986)]{AHMR[1986]}
Abarbanel H.D.I., Holm D. D., Marsden J. E. and Ratiu T.,
Nonlinear stability analysis of stratified ideal fluid equilibria.
{\it Phil Trans. Roy. Soc.}, (London) A {\bf 318} (1986) 349--409.

\bibitem[Allen and Holm(1996)]{AH[1996]} Allen J. S. and Holm D. D.,
Extended-geostrophic Hamiltonian models for rotating
shallow water motion.
{\it Physica D}, {\bf 98} (1996) 229--248.

\bibitem[Allen, Holm and Newberger(1998)]{AHN[1998]} 
Allen J. S., Holm D. D. and Newberger P.,
Toward an extended-geostrophic Euler--Poincar\'e
model for mesoscale oceanographic flow.
In {\it Proceedings of the Isaac Newton Institute Programme on
the Mathematics of Atmospheric and Ocean Dynamics}, 
(Cambridge University Press, Cambridge) to appear.

\bibitem[Bannon(1995)]{Bannon[1995]} Bannon P. R., 
Hydrostatic adjustment: Lamb's problem.
{\it J. Atmos. Sci.}, {\bf 52} (1995) 1743-1752.

\bibitem[Bannon(1996)]{Bannon[1996]} Bannon P. R.,
On the anelastic approximation for a compressible atmosphere.
{\it J. Atmos. Sci.}, {\bf 53} (1996) 3618-3628.

\bibitem[Bernardet(1995)]{Bernardet[1995]} Bernardet P., 
The pressure term in the anelastic model: a symmetric elliptic solver
for an Arakawa C grid in generalized coordinates.
{\it Monthly Weather Rev.}, {\bf 123} (1995) 2474-2490.

\bibitem[Bloch et al.(1996)]{BKMM[1996]}
Bloch A. M., Krishnaprasad P. S., Marsden J. E. and Murray R., 
Nonholonomic mechanical systems with symmetry.
{\it Arch. Rat. Mech. An.}, {\bf 136} (1996) 21--99.

\bibitem[Cendra et al.(1987)]{Cendra.etal[1987]}
Cendra H., Ibort A. and Marsden J. E., 
Variational principal fiber bundles: a geometric
theory of Clebsch potentials and Lin constraints. 
{\it J.  Geom.  Phys.\/}, {\bf 4} (1987) 183--206.

\bibitem[Cendra et al.(1998a)]{Cendra.etal[1998a]}
Cendra H., Holm D. D., Marsden J. E. and Ratiu T. S.,  
Lagrangian reduction, the Euler--Poincar\'{e}
equations, and  semidirect products. 
In {\it Arnol'd Festschrift Volume II}, 
{\bf186} (1998) Am. Math. Soc. Translations Series 2,
pp 1-25. 

\bibitem[Cendra et al.(1998b)]{Cendra.etal[1998b]}  
Cendra H., Marsden J. E. and Ratiu T. S., 
Lagrangian reduction by stages. (1998) 
{\it In preparation.}

\bibitem[Durran(1989)]{Durran[1989]} Durran D. R.,
Improving the anelastic approximation.
{\it J. Atmos. Sci.}, {\bf 46} (1989) 1453-1461.

\bibitem[Dutton and Fichtl(1969)]{DF[1969]} Dutton D. R. and Fichtl G. H.,
Approximate equations of motion for gases and liquids.
{\it J. Atmos. Sci.}, {\bf26} (1969) 241-254.

\bibitem[Holm(1996)]{Holm[1996]} Holm D. D.,
Hamiltonian balance equations.
{\it Physica D}, {\bf 98} (1996) 379--414.

\bibitem[Holm, Marsden and Ratiu(1986)]{HMR[1986]}
Holm D. D., Marsden J. E. and Ratiu T.,
The Hamiltonian Structure of 
Continuum Mechanics in Material, Inverse
Material, Spatial and Convective Representations,  
in {\it Hamiltonian Structure and
Lyapunov Stability for Ideal Continuum  Dynamics}, 
by D. D. Holm, J. E. Marsden, and T. S. Ratiu, 
(Univ. Montreal Press, Montreal) 1986, pp. 1-124.

\bibitem[Holm, Marsden and Ratiu(1998a)]{HMR[1998a]}  
Holm D. D., Marsden J. E. and Ratiu T.,
The Euler-Poincar\'e equations and semidirect
products with applications to continuum theories, 
{\it Adv. in Math.}, {\bf 137} (1998) 1-81.

\bibitem[Holm, Marsden and Ratiu(1998b)]{HMR[1998b]}  Holm D. D., Marsden J. E. and Ratiu T.,
The Euler--Poincar\'{e} equations in geophysical fluid dynamics,
in {\it Proceedings of the Isaac Newton Institute Programme on
the Mathematics of Atmospheric and Ocean Dynamics}, 
(Cambridge University Press, Cambridge) to appear.

\bibitem[Holm, Marsden and Ratiu(1998c)]{HMR[1998c]}  
Holm D. D., Marsden J. E. and Ratiu T., 
Euler-Poincare models of ideal fluids with nonlinear dispersion. 
{\it Phys. Rev. Lett.}, {\bf 80} (1998) 4173--4177. 

\bibitem[Holm and Zeitlin(1998)]{HZ[1998]} Holm D. D. and Zeitlin V.,
Hamilton's principle for quasigeostrophic motion.
{\it Phys. Fluids}, {\bf 10} (1998) 800-806.

\bibitem[Lilly(1994)]{Lilly[1994]} Lilly D. K.,
A comparison of incompressible, anelastic and Boussinesq dynamics. 
{\it Atmosph. Res.}, {\bf 40} (1994) 143-151.

\bibitem[Lipps and Hemler(1982)]{LH[1982]} Lipps F. B. and Hemler R. S., 
A scale analysis of deep moist convection and some related numerical
calculations. {\it J. Atmos. Sci.}, {\bf39} (1982) 2192-2210.

\bibitem[Marsden and Ratiu(1994)]{MR[1994]}
Marsden J. E. and Ratiu T. S.,  {\it Introduction to
Mechanics and Symmetry.\/} Texts in Applied Mathematics, {\bf  17},
Springer-Verlag 1994.

\bibitem[Marsden and Scheurle(1993a)]{MS[1993a]} 
Marsden J. E. and Scheurle J., 
Lagrangian reduction and the double spherical pendulum.
{\it ZAMP\/}, {\bf 44} (1993) 17--43.

\bibitem[Marsden and Scheurle(1993b)]{MS[1993b]}
Marsden J. E. and Scheurle J., 
The reduced Euler-Lagrange equations.
{\it Fields Institute Comm.\/}, {\bf 1} (1993) 139--164.

\bibitem[Marsden et al.(1998)]{MPS[1998]} 
Marsden J. E., Patrick G. W. and Shkoller S., 
Multisymplectic geometry, variational integrators,
and nonlinear PDEs. (1998) {\it Comm. Math. Physics},
to appear.

\bibitem[Nance and Durran(1994)]{ND[1994]} 
Nance L. B. and Durran D. R.,
A comparison of 3 anelastic systems and the pseudo-incompressible system.
{\it J. Atmos. Sci.}, {\bf 51} (1994) 3549-3565.

\bibitem[Scinocca and Shepherd(1992)]{SS[1992]} Scinocca J. F. and Shepherd T. G., 
Nonlinear wave-activity conservation laws and Hamiltonian structure
for the two-dimensional anelastic equations. 
{\it J. Atmos. Sci.}, {\bf 49} (1992) 5-27.



\end{thebibliography}

\end{document}